%Paper: hep-th/9303104
%From: Per Berglund <BERGLUND@sbitp.itp.ucsb.edu>
%Date: Thu, 18 Mar 1993 13:14 PST

%%%%%%%%%%%%%%%%%%%%%%%%%%%%%%%%%%%%%%%%%%%%%%%%%%%%%%%%%%%%%%%%%%%%%%%%%%%
%%                                                                       %%
%%   HDMDT.TEX   by    Per Berglund                                      %%
%%                  berglund@sbitp.ucsb.edu                              %%
%%%%%%%%%%%%%%%%%%%%%%%%%%%%%%%%%%%%%%%%%%%%%%%%%%%%%%%%%%%%%%%%%%%%%%%%%%%
\input harvmac
%%%%%%%%%%%%%%%%%%%%%%%%%%%%%%%%%%%%%%%%%%%%%%%%%%%%%%%%%%%%%%%%%%%%%%%%
%                                                                      %
%       "zip.tex", a set of macros to be used with "harvmac.tex"       %
%            Latest change: 27. VII '92.   (Tristan Hubsch)            %
%                                                                      %
%%%%%%%%%%%%%%%%%%%%%%%%%%%%%%%%%%%%%%%%%%%%%%%%%%%%%%%%%%%%%%%%%%%%%%%%
 %
\catcode`@=11
\def\rlx{\relax\leavevmode}                  % Guess what this is for...
 %
 %
%%%%%%%%%%%%%%%%%%%%%%%%%%%%%%%%%%%%%%%%%%%%%%%%%%%%%%%%%%%%%%%%%%%%%%%%
%%%****FONTS****FONTS****FONTS****FONTS****FONTS****FONTS****FONTS****%%
 % That is, where fonts may not be available...
 %
 % Poor man's boldface; use in lieu of proper boldface
\def\BM#1{\relax\leavevmode\setbox0=\hbox{$#1$}
           \kern-.025em\copy0\kern-\wd0
            \kern.05em\copy0\kern-\wd0
             \kern-.025em\raise.0433em\copy0\kern-\wd0
              \raise.0144em\box0 }
 %
 % Some basic black-board bold (capital) letters
 % ...should work rather well in sub- and super-scripts also...
 %
\def\inbar{\vrule height1.5ex width.4pt depth0pt}
\def\sinbar{\vrule height1ex width.35pt depth0pt}
\def\ssinbar{\vrule height.7ex width.3pt depth0pt}
\font\cmss=cmss10
\font\cmsss=cmss10 at 7pt
\def\ZZ{\rlx\leavevmode
             \ifmmode\mathchoice
                    {\hbox{\cmss Z\kern-.4em Z}}
                    {\hbox{\cmss Z\kern-.4em Z}}
                    {\lower.9pt\hbox{\cmsss Z\kern-.36em Z}}
                    {\lower1.2pt\hbox{\cmsss Z\kern-.36em Z}}
               \else{\cmss Z\kern-.4em Z}\fi}
\def\Ik{\rlx{\rm I\kern-.18em k}}  % Yes, I know. This ain't capital.
\def\IC{\rlx\leavevmode
             \ifmmode\mathchoice
                    {\hbox{\kern.33em\inbar\kern-.3em{\rm C}}}
                    {\hbox{\kern.33em\inbar\kern-.3em{\rm C}}}
                    {\hbox{\kern.28em\sinbar\kern-.25em{\sevenrm C}}}
                    {\hbox{\kern.25em\ssinbar\kern-.22em{\fiverm C}}}
             \else{\hbox{\kern.3em\inbar\kern-.3em{\rm C}}}\fi}
\def\IP{\rlx{\rm I\kern-.18em P}}
\def\IR{\rlx{\rm I\kern-.18em R}}
\def\Ione{\rlx{\rm 1\kern-2.7pt l}}
 %
%%%%%%%%%%%%%%%%%%%%%%%%%%%%%%%%%%%%%%%%%%%%%%%%%%%%%%%%%%%%%%%%%%%%%%%%
%%%****SHAPE****SHAPE****SHAPE****SHAPE****SHAPE****SHAPE****SHAPE****%%
 %
 % Get in shape, Man, (never mind the content)!
\def\boxit#1{\vbox{\hrule\hbox{\vrule\kern3pt
     \vbox{\kern3pt#1\kern3pt}\kern3pt\vrule}\hrule}}

\def\intem#1{\par\leavevmode%
              \llap{\hbox to\parindent{\hss{#1}\hfill~}}\ignorespaces}
 %
 % Indents #1 lines by width of #2 and puts #2 in the "niche".

 % Similar to "niche" :

 %
 % Math...
 % My version of \eqalign, \eqalignno ...
\newskip\humongous \humongous=0pt plus 1000pt minus 1000pt   % isn't it?
\def\caja{\mathsurround=0pt}
\newif\ifdtup
 %
 % display pattern: [         a &= b        \cr]
\def\eqalign#1{\,\vcenter{\openup2\jot \caja
     \ialign{\strut \hfil$\displaystyle{##}$&$
      \displaystyle{{}##}$\hfil\crcr#1\crcr}}\,}
 %
 % display pattern: [   a &= b  &  c &= d   \cr]
\def\twoeqsalign#1{\,\vcenter{\openup2\jot \caja
     \ialign{\strut \hfil$\displaystyle{##}$&$
      \displaystyle{{}##}$\hfil&\hfill$\displaystyle{##}$&$
       \displaystyle{{}##}$\hfil\crcr#1\crcr}}\,}
 %
 % display to full hsize, numbered at far right
\def\panorama{\global\dtuptrue \openup2\jot \caja
     \everycr{\noalign{\ifdtup \global\dtupfalse
      \vskip-\lineskiplimit \vskip\normallineskiplimit
      \else \penalty\interdisplaylinepenalty \fi}}}
 %
 % display pattern: [         a &= b        &(*) \cr]
\def\eqalignno#1{\panorama \tabskip=\humongous
     \halign to\displaywidth{\hfil$\displaystyle{##}$
      \tabskip=0pt&$\displaystyle{{}##}$\hfil
       \tabskip=\humongous&\llap{$##$}\tabskip=0pt\crcr#1\crcr}}
 %
 % display pattern: [      a &= b &= c      &(*) \cr]

 %
 % display pattern: [   a &= b  &  c &= d   &(*) \cr]

 %
 % display pattern: [    a &= b &= c &= d   &(*) \cr]

 %
 % For extra v-space between rows of a matrix or eqn-alignment,
 % use "\noalign{\vskip2mm}". In the above equation alignments,
 % "\openup2mm" does the same, but has no effect in "\matrix".
 %
%%%%%%%%%%%%%%%%%%%%%%%%%%%%%%%%%%%%%%%%%%%%%%%%%%%%%%%%%%%%%%%%%%%%%%%%
%%%****SHORT****SHORT****SHORT****SHORT****SHORT****SHORT****SHORT****%%
 %
 % Redefinitions of TeX's commands :
 %
          % Polish l-slash, L-slash
        % Scandinavian o-slash, O-slash
          % P-mirror, double-S
        % tie-after, cedilla
          % include mathmode !!!
          % under-bar, under-dot
\def\,{\hskip1.5pt}           % why only in math-mode?
 %
 % Some abbreviations that save typing :
\let\a=\alpha

\let\c=\chi
                    
\let\e=\epsilon

\let\l=\lambda                                   
\let\m=\mu
\let\n=\nu
\let\p=\pi                         
\let\q=\theta                   \let\Q=\Theta
         
\let\s=\sigma

 %
 % Additional math symbols
 %
\def\Box{\sqcap\llap{$\sqcup$}}
\def\lapp{\lower.4ex\hbox{\rlap{$\sim$}} \raise.4ex\hbox{$<$}}
\def\gapp{\lower.4ex\hbox{\rlap{$\sim$}} \raise.4ex\hbox{$>$}}
\def\con{\ifmmode\raise.1ex\hbox{\bf*}
          \else\raise.1ex\hbox{\bf*}\fi}
\def\bo{{\raise.15ex\hbox{\large$\Box\kern-.39em$}}}
  %  a very fat nothing

\def\dual{\relax\leavevmode\lower.9ex\hbox{\titlerms*}}
\def\define{\buildrel\rm def\over =}

\let\8=\otimes
 %
 %
%%%%%%%%%%%%%%%%%%%%%%%%%%%%%%%%%%%%%%%%%%%%%%%%%%%%%%%%%%%%%%%%%%%%%%%%
%%****MACROS***MACROS***MACROS***MACROS***MACROS***MACROS***MACROS****%%
 %
 % Math macros
 %

\let\2=\underline

 %
 % Let's take arguments...
%  Use:  "A\like{B}".
\def\dt#1{{\buildrel{\smash{\lower1pt\hbox{.}}}\over{#1}}}

\font\eightrm=cmr8
\def\6(#1){\relax\leavevmode\hbox{\eightrm(}#1\hbox{\eightrm)}}
\def\0#1{\relax\ifmmode\mathaccent"7017{#1}     % a little circle atop,
                \else\accent23#1\relax\fi}      % as a halo of a saint
\def\7#1#2{{\mathop{\null#2}\limits^{#1}}}      % puts #1 atop #2
\def\5#1#2{{\mathop{\null#2}\limits_{#1}}}      % puts #1 beneath #2
 %
 % Will grow vertically with size of argument

 %
 % Vertical arrows with labels

 %
 % For vert. arrows to grow, say "\bigg\down\crlap{...}", using the
 % side-script: a label for vertical delimiters

 %

 %
 % Horizontal arrows that can grow
\newbox\t@b@x
\def\rightarrowfill{$\m@th \mathord- \mkern-6mu
     \cleaders\hbox{$\mkern-2mu \mathord- \mkern-2mu$}\hfill
      \mkern-6mu \mathord\rightarrow$}
\def\tooo#1{\setbox\t@b@x=\hbox{$\scriptstyle#1$}%
             \mathrel{\mathop{\hbox to\wd\t@b@x{\rightarrowfill}}%
              \limits^{#1}}\,}
\def\leftarrowfill{$\m@th \mathord\leftarrow \mkern-6mu
     \cleaders\hbox{$\mkern-2mu \mathord- \mkern-2mu$}\hfill
      \mkern-6mu \mathord-$}
\def\froo#1{\setbox\t@b@x=\hbox{$\scriptstyle#1$}%
             \mathrel{\mathop{\hbox to\wd\t@b@x{\leftarrowfill}}%
              \limits^{#1}}\,}
 %
 % fractions
\def\frac#1#2{{#1\over#2}}
\def\frc#1#2{\relax\ifmmode{\textstyle{#1\over#2}} % A small fraction,
                    \else$#1\over#2$\fi}           % good in text.
                            % Like {1\over{#1}}
 %
 % The basic Theorem-like macro, uses equation numbers
 % Use:  \Claim\cLABEL{Theorem}{This is a theorem.}
\def\Claim#1#2#3{\bigskip\begingroup%
                  \xdef #1{\secsym\the\meqno}%
                   \writedef{#1\leftbracket#1}%
                    \global\advance\meqno by1\wrlabeL#1%
                     \noindent{\bf#2}\,#1{}\,:~\sl#3\vskip1mm\endgroup}

\def\QED{\rlx\hfill$\Box$\kern-7pt\raise3pt\hbox{$\surd$}\bigskip}
 %
 % Math miscellanea
 %
\def\Tr{\mathop{\rm Tr}}

\def\WK#1#2{\relax\def\normalbaselines{\baselineskip12pt\lineskip3pt
                                       \lineskiplimit3pt}
              \matrix{#1}\left[\,\matrix{#2}\right]}
\def\muthstrut{\vphantom1}
\def\mutrix#1{\null\,\vcenter{\normalbaselines\m@th
        \ialign{\hfil$##$\hfil&&~\hfil$##$\hfill\crcr
            \muthstrut\crcr\noalign{\kern-\baselineskip}
            #1\crcr\muthstrut\crcr\noalign{\kern-\baselineskip}}}\,}

 %
 % Young tableaux: use "\Box" for a box and "\Z" for newline
 % The 2-1 stair-tableau:  \YT{\Box\Box}{\Box\Box\Z\Box}
 % Note: the first argument sets the width!
\def\YT#1#2{\vcenter{\hbox{\vbox{\baselineskip=\normalbaselineskip%
             \def\Box{$\sqcap\llap{$\sqcup$}$\kern-1.2pt}%
              \def\Z{\hfil\vskip-8.8pt}%
               \setbox0=\hbox{#1}\hsize\wd0\parindent=0pt#2}\,}}}
\def\EU{\rlx\ifmmode \c_{{}_E} \else$\c_{{}_E}$\fi}
\def\TM{\rlx\ifmmode {\cal T_M} \else$\cal T_M$\fi}
\def\TW{\rlx\ifmmode {\cal T_W} \else$\cal T_W$\fi}
\def\CM{\rlx\ifmmode {\cal T\rlap{\bf*}\!\!_M}
             \else$\cal T\rlap{\bf*}\!\!_M$\fi}
\def\hm#1#2{\rlx\ifmmode H^{#1}({\cal M},{#2})
                 \else$H^{#1}({\cal M},{#2})$\fi}
\def\CP#1{\rlx\ifmmode\IP^{#1}\else\IP$^{#1}$\fi}
\def\cP#1{\rlx\ifmmode\IC{\rm P}^{#1}\else$\IC{\rm P}^{#1}$\fi}
\def\WCP#1#2{\rlx\IP^{#1}_{#2}}
\def\sll#1{\rlx\rlap{\,\raise1pt\hbox{/}}{#1}}
\def\Sll#1{\rlx\rlap{\,\kern.6pt\raise1pt\hbox{/}}{#1}\kern-.6pt}
%

 %
 % Text miscellanea
 %
\def\ie{\hbox{\it i.e.}}        % By Knuth, use commas: ..., \ie, ... !

\def\CY{Calabi-\kern-.2em Yau}
\def\LG{Landau-Ginzburg}
\def\LGO{Landau-Ginzburg orbifold}
\def\3{\ifmmode\ldots\else$\ldots$\fi}
\def\Z{\hfil\break\rlx\hbox{}\quad}
\def\3{\ifmmode\ldots\else$\ldots$\fi}
\def\?{d\kern-.3em\raise.64ex\hbox{-}}           % d-dash
\def\9{\raise.43ex\hbox{-}\kern-.37em D}         % D-Dash

 %
 % References
 %

 %

\def\pre#1{{\it University of #1 report}}

\def\NP#1{{\it Nucl.\,Phys.\,}{\bf#1\,}}
\def\PL#1{{\it Phys.\,Lett.\,}{\bf#1\,}}

\def\CMP#1{{\it Commun.\,Math.\,Phys.\,}{\bf#1\,}}

 %
 %
%%%%%%%%%%%%%%%%%%%%%%%%%%%%%%%%%%%%%%%%%%%%%%%%%%%%%%%%%%%%%%%%%%%%%%%%
%%============>>>>            SAVE  TIMBER            <<<<============%%
 %
\baselineskip=13.0861pt plus2pt minus1pt
\parskip=\medskipamount
\let\ft=\foot
\noblackbox
\def\SaveTimber{\abovedisplayskip=1.5ex plus.3ex minus.5ex
                \belowdisplayskip=1.5ex plus.3ex minus.5ex
                \abovedisplayshortskip=.2ex plus.2ex minus.4ex
                \belowdisplayshortskip=1.5ex plus.2ex minus.4ex
                \baselineskip=12pt plus1pt minus.5pt
 \parskip=\smallskipamount
 \def\ft##1{\unskip\,\begingroup\footskip9pt plus1pt minus1pt\setbox%
             \strutbox=\hbox{\vrule height6pt depth4.5pt width0pt}%
              \global\advance\ftno by1\footnote{$^{\the\ftno)}$}{##1}%
               \endgroup}
 \def\listrefs{\footatend\vfill\immediate\closeout\rfile%
                \writestoppt\baselineskip=10pt%
                 \centerline{{\bf References}}%
                  \bigskip{\frenchspacing\parindent=20pt\escapechar=` %
                   \rightskip=0pt plus4em\spaceskip=.3333em%
                    \input refs.tmp\vfill\eject}\nonfrenchspacing}}
 %
 % For European standard
\def\Afour{\ifx\answ\bigans
            \hsize=16.5truecm\vsize=24.7truecm
             \else
              \hsize=24.7truecm\vsize=16.5truecm
               \fi}
\catcode`@=12
%%%%%%%%%%%%%%%%%%%%%%%%%%%%%%%%%%%%%%%%%%%%%%%%%%%%%%%%%%%%%%%%%%%%%%%%%
%                                                                       %
%                          End of  "zip.tex".                           %
%                        May the article begin!                         %
%                                                                       %
%%%%%%%%%%%%%%%%%%%%%%%%%%%%%%%%%%%%%%%%%%%%%%%%%%%%%%%%%%%%%%%%%%%%%%%%%

\def\cp#1#2{\hbox{$\IP_{#1}^{#2}$}}
\def\eqaligntwo#1{\,\vcenter{\openup2\jot \caja
     \ialign{\strut \hfil$\displaystyle{##}$&
                         $\displaystyle{{}##}$\hfil&
                          $\displaystyle{{}##}$\hfil\crcr#1\crcr}}\,}

 \def\Afour{\hsize=16.5truecm\vsize=24.7truecm}
% \Afour
\SaveTimber

\Title{\vbox{\baselineskip12pt \hbox{NSF-ITP-93-27}
                               \hbox{UTTG-07-93}}}
      {\vbox{\centerline{Dimensionally Reduced Landau-Ginzburg Orbifolds}
             \vskip10pt
             \centerline{ with
                         Discrete Torsion}}}

\centerline{Per Berglund\footnote{$^{\diamondsuit}$}
            {After Sept.~15th: Institute for Advanced Study,
             Olden Lane, Princeton, NJ 08540.}}
                                                    \vskip0mm
 \centerline{Institute for Theoretical Physics}      \vskip-1mm
 \centerline{University of California}               \vskip-1mm
 \centerline{Santa Barbara, CA 93106}                \vskip0mm
 \centerline{and}                                    \vskip0mm
 \centerline{Theory Group, Department of Physics}    \vskip-1mm
 \centerline{University of Texas}    \vskip-1mm
 \centerline{Austin, TX 78712}   \vskip-1mm
 \centerline{berglund@sbitp.ucsb.edu} \vskip0mm
 \vfill

\centerline{ABSTRACT}\vskip2mm
\vbox{\narrower\narrower\baselineskip=12pt
It is observed that a large class of $(2,2)$ string vacua with
$n>5$ superfields
can be rewritten as Landau-Ginzburg orbifolds with discrete
torsion and $n=5$. The naive geometric interpretation (if
one exists)  would be that of a complex $3$-fold, not necessarily
K\"ahler but still with vanishing first Chern class.}

\Date{\vbox{
%      \line{CERN-TH-6???/92\hfill}
       \line{3/\number\yearltd \hfill}}}

\noblackbox

\nref\rDixon{For a review and references, see L.~Dixon: in {\it
     Superstrings, Unified Theories and Cosmology 1987},\,p.67--127,
     eds.~G.~Furlan et al.\ (World Scientific, Singapore, 1988).}
\nref\rCHSW{P.~Candelas, G.~Horowitz, A.~Strominger and E.~Witten~:
      \NP{B258}(1985)46.}
\nref\rMPR{P.~Candelas, M.~Lynker and R.~Schimmrigk:
      \NP{B341}(1990)383\semi
B.~R.~Greene and M.~R.~Plesser: \NP{B338}(1990)15\semi
P.~S.~Aspinwall, C.~A.~L\"utken and G.~G.~Ross:
      \PL{241B}(1990)373.}
\nref\rPLE{P.~Candelas, L.~Parkes and E.~Derrick: ``Generalized \CY\
Manifolds and the Mirror of a Rigid Manifold'',
\pre{Texas} UTTG-24-92.}
\nref\rR{R.~Schimmrigk~: ``Critical Superstring Vacua from Noncritical
Manifolds: A Novel Framework'',
      {\it Universit\"at Heidelberg report} HD-THEP-92-29.}
\nref\rVMS{C.~Vafa~: ``Toplogical Mirrors and Quantum Rings'' in
     {\it ``Essays on Mirror Manifolds"}, S.T.~Yau ed.
      International Press, Hong Kong 1992.}
\nref\rS{A.~Strominger~: \NP{B274} (1986)253.}
\nref\rLS{M.~Lynker and R.~Schimmrigk: \PL{249B}(1990)237.}
\nref\rVW{C.~Vafa and N.P.~Warner: \PL{218B} (1989) 377\semi
          E.Martinec: \PL{B217} (1989) 431.}
%\nref\rVS{C.~Vafa: \MPL{A4} (1989) 1169.}
\nref\rAR{A.~Klemm and R.~Schimmrigk~: ``Landau-Ginzburg String Vacua'',
{\it CERN preprint} CERN-TH 6459/92\semi
          M.~Kreuzer and H.~Skarke: ``No Mirror Symmetry in Landau-Ginzburg
 Spectra'', {\it CERN preprint} CERN-TH 6461/92.}
\nref\rGV{B.R.~Greene and C.~Vafa: unpublished.}
\nref\rBH{P.~Berglund and T.~H\"ubsch: ``A Generalized Construction
of Mirror Manifolds'' in  {\it ``Essays on Mirror Manifolds"}, S.T.~Yau ed.
      International Press, Hong Kong 1992; also to appear in \NP{B}.}
\nref\rIV{K.~Intrilligator and C.~Vafa: \NP{B339} (1990) 95.}
\nref\rFIQ{A.~Font, L.E.~Ib\'a\~nez and F.~Quevedo:\PL{B217}(1989) 272.}
\nref\rVT{C.~Vafa: \NP{B273}(1986)592.}
\nref\rFK{J.~Fuchs, A.~Klemm, C.~Scheich and M.G.~Schmidt:
{\it Ann. Phys.} {\bf 204} (1990) 1.}
\nref\rGH{P.~Green and T.~H\"ubsch: \CMP{113} (1987) 505.}
\nref\rGVW{B.R.~Greene, C.~Vafa and N.P.~Warner:
     \NP{B324}(1989)371.}
\nref\rROLF{R.~Schimmrigk: \PL{193B} (1987)175.}
\nref\rRS{R.~Schimmrigk: \PL{229B} (1989) 227.}
\nref\rTEX{P.~Candelas, A.M.~Dale, C.A.~L\"utken and R.~Schimmrigk:
     \NP{B298}(1988)493.}

\newsec{Introduction}
In compactifying the ten dimensional heterotic string theory to our
physical four dimensional space-time the extra degrees of freedom have
to form an $N=2$ superconformal field theory (SCFT) with central
charge $c=9$ \rDixon\ .
(To be more precise,
$(2,0)$ SCFT is sufficient for $N=1$ space time supersymmetry.  However, in our
discussion we will restrict our attention to the left-right symmetric
$(2,2)$ models.)

One way of realizing the internal theory is to let it
be described by a three dimensional K\"ahler manifold with vanishing
first Chern class, \ie\  a Calabi-Yau manifold \rCHSW\ . In the light
of mirror symmetry \rMPR\ this gives rise to the
following problem.
In terms of a $(2,2)$ SCFT mirror symmetry is the phenomenon that
by changing the sign of the left-moving $U(1)$-current we will get a
new theory which is isomorphic to the original one while interchanging the
$(1,1)$ and $(-1,1)$ states. In the geometric picture (when such
exists), and restricting to an odd (complex) dimensional  target space,
 the corresponding manifolds have  opposite Euler
number and the number of complex deformations (denoted by $b_{2,1}$)
and K\"ahler deformations ($b_{1,1}$) interchanged\footnote{$^1$}
{We adopt the {\it
convention}~\rDixon\ where $(2,1)$-forms are equivalent to $U(1)$
charge-$(1,1)$ states in the (chiral,chiral) sector while $(1,1)$-forms are
analogs of charge-$(-1,1)$ states in the (anti-chiral, chiral) sector.}.
 The existence of rigid manifolds, $b_{2,1}=0$, implies that mirror
symmetry would take us outside the class of \CY\ three folds since
the mirror would not be K\"ahler. It turns out that a possible resolution to
this problem is
to consider some higher dimensional manifold, a hypersurface in a
weighted projective space of dimension larger than four \refs{\rPLE,\rR}.
The observation is that, at least for some particular examples,
the middle cohomology class is the mirror symmetric one with respect
to the original rigid manifold.

However this leads to another problem: How are we to think of these
theories as target spaces for a non-linear $\s$-model? The
simplest resolution is just to say that we cannot always associate a
classical geometry to a $(2,2)$ SCFT. Another way out of this
dilemma   is to say that the
higher dimensional (non-critical) manifold contains the mirror of the rigid
manifold as a submanifold \rR\ . A third possibility would be to
introduce non-trivial torsion and to turn on the dilaton field \rVMS\ ;
these three dimensional complex, non K\"ahler manifolds are
consistent solutions to the classical string equations \rS\ .

In this paper we will take a first step in the latter direction.
We start by considering an orbifold of a  Landau-Ginzburg potential
with $n>5$ superfields.
The
key ingredient in the construction that we propose is a $\ZZ_2$ field
redefinition,
\ie\ a fractional transformation \rLS\ , which allows us to rewrite the
superpotential such that trivial (quadratic) terms can be neglected at
the expense of introducing $\ZZ_2$ discrete torsion.
It turns out that this
technique applies to a very large class of $(2,2)$ string vacua
for which the theory can be written as a dimensionally reduced ($n=5$)
Landau-Ginzburg orbifold with non-trivial discrete torsion.
It is tempting to
conjecture that the geometric interpretation of these reduced Landau-Ginzburg
orbifolds with discrete torsion indeed would be that of   three-folds
of the type described by Strominger \rS\ .

The paper is organized as follows. We start by describing the
construction in terms of  the Landau-Ginzburg formalism (section 2).
In section 3 we discuss the limitations of our analysis  while our conclusions
are left for section 4.

%In this note we point out a possible solution to the problem of
%the mirror of rigid manifolds or in more general to manifolds without
%K\"ahler class. This is related to having higher dimensional manifolds
%as string vacua and the problem of formulating a $\s$-model with
%these manifolds as target spaces.

\newsec{The Construction}

Consider the usual action of an $N=2$ superconformal Landau-Ginzburg theory
\eqn\eACTION{\int d^2z d^2\q d^2\bar\q K(X_i,\bar X_i) +
             (\int d^2z d^2\q W(X_i) + c.c).}
where $K$ is the K\"ahler potential and $W$, the superpotential, is a
holomorphic function of the $N=2$ chiral superfields
$X_i(z,\bar z,\q^+,\q^-)$. Due to
nonrenormalization theorems  $W$ is not renormalized (up to scaling)
and hence will
characterize the theory (modulo irrelevant perturbations coming from
the K\"ahler potential \rVW\ .) Let $W$ be a polynomial in the  superfields
$X_i~,~i=1,\ldots,n$. We require that $W$ is quasi-homogenous of degree $d$,
 \ie\  under rescaling of the world-sheet
\eqn\eSS{ X_i \mapsto \l^{k_i} X_i~~,\qquad W(X_i) \mapsto \l^d W(X_i)~.}
 To compute the central charge is straight forward \rVW\ ,
 $c=6\sum_{i=1}^n
({1\over 2} - q_i)$ with $q_i=k_i/d$ the charge under the $U(1)~$-
current, $J_0$.
In order to have $(2,0)$ world-sheet supersymmetry we need to consider
the LG orbifold $W(X_i)/j$ where $j=e^{2\p iJ_0}$. This projects onto
the integer charged states and preserves the supersymmetry.
We will restrict our attention to models with $c=9$ and those
 with $n=5,7,9$.
\footnote{$^2$}{For a recent complete construction of such
Landau-Ginzburg string
vacua see Ref. \rAR\ .}
 This is equivalent to the statement that
\eqn\eCCC{
\sum_{i=1}^n k_i = kd
}
where $k=(n-3)/2$.

Let $n=7$ and define ${\cal M}=W(X_i)/j$ where we choose $W(X_i)$ to
 be a polynomial of the form considered
in \rBH\ . For these theories it is possible to write a potential with only
$n$ terms such that the origin is the only singular point.
Furthermore we restrict $W(X_i)$ to be either of the two below
\eqna\eWGEN
$$\eqalignno{W(X_i)~~&=~~W_0(X_i) + X_4 X_6^2 + X_5 X_7^2~, & \eWGEN a \cr
             W(X_i)~~&=~~W_0(X_i) + X_4 X_6^2 + X_7^2~. & \eWGEN b \cr}$$
where $W_0(X_i)$ is of even degree and
does not contain the fields $X_6$ or $X_7$.
{}From now on we will only consider Eq. \eWGEN a\ ; \eWGEN b\
follows in a similar fashion.
Note that due to
the particular form of $W(X_i)$,  $X_4$ appears only in one term of
$W_0(X_i)$, either as $X_4^l$ or as $X_4^lX_i$ for some field $X_i$.
The same is true for $X_5$.
We then
make the following field redefinitions (with constant Jacobian)
\footnote{$^3$}{The unpaired zero-modes of the $X_i$
(the number of which is $\Tr(-1)^F$) gives
a non-trivial piece to the (super) Jacobian while the other zero (and higher)
modes do not contribute due to supersymmetry \rGV\ . Hence
 the Jacobian is {\it not} trivially a constant .}
\eqn\eREDEF{
\twoeqsalign{\tilde X_4~&=~X_4^{1/2}~,
                      &~~\tilde X_6~&=~X_6 X_4^{1/2}~,  \cr
                 \tilde X_5~&=~X_5^{1/2}~,
                      &~~\tilde X_7~&=~X_7 X_5^{1/2}~,  \cr}
}
and $\tilde X_i=X_i~,~i=1,2,3~.$  In order to make this a well defined
map we need to impose the $\ZZ_2^2$ identifications on the $\tilde X_i$,
corresponding to a quotient by
\footnote{$^3$} {We will use
the notation $(\ZZ_k:\Q_1,\Q_2,\Q_3,\Q_4,\Q_5,\Q_6,\Q_7)$ for a $\ZZ_k$
symmetry
with the action\hfill\break $(X_1,X_2,X_3,X_4,X_5,X_6,X_7) \to (\a^{\Q_1}
X_1,\ldots,\a^{\Q_7} X_7)$, where
$\a^k~=~1$.}
\eqn\eIDENT{
\twoeqsalign{h_1~~&\define~~(\ZZ_2:0,0,0,1,0,1,0)~, &
             \qquad h_2~~&\define~~(\ZZ_2:0,0,0,0,1,0,1)~. \cr}
}
Thus ${\cal M}\simeq \widetilde W(\tilde X_i)/(\tilde j\cdot h_1\cdot h_2)$
where
\eqn\eWREDEF{
\widetilde W(\tilde X_i)~~=~~
\widetilde W_0(\tilde X_i) + \tilde X_6^2 + \tilde X_7^2~,}
and $\tilde j$ is the new scaling symmetry.
Up to this point we have merely rewritten the original theory. The
crucial observation is that we can neglect the quadratic piece at the
expense of introducing non-trivial torsion. Thus we would
get a LG orbifold with $n=5$, which naively corresponds to a
manifold with trivial canonical class (see \eCCC\ ). The question is how one
is to interpret the torsion in terms of the geometric
picture. We will return to this discussion in section 4. For now
let us restrict the analysis to the LG orbifold.

In order to give the argument we need to set up some notation
along the lines of Ref. \rIV\ .
Let us consider the action of a group element $g\in Q$, where $Q$ is the
quotient group with respect to which the orbifold is constructed,
on a physical state in the $h$ twisted $(c,c)$
sector \rIV\ ,
\eqn\eGACT{
\eqalign{g & \prod_{\Q_i^h\in\ZZ}  (X_i)^{\ell_i}
\Big|0\Big\rangle^{(h)}_{(c,c)}~ = \cr
\phantom{g }~&=\e(g,h)(-1)^{K_gK_h}(\det g)^{-1}
\det g|_h \exp\left(2\p i\sum_{\Q_i^h\in\ZZ} \Q_i^g\ell_i\right)
\prod_{\Q_i^h\in\ZZ} (X_i)^{\ell_i} \Big|0\Big\rangle^{(h)}_{(c,c)}~,}
}
where $\Q_i^h$ is the twist charge for the $i$th field under the $h$-twist
 and $\det g|_h$ is the determinant of $g$ restricted to
the fields $X_i$ left untwisted by $h$.
The different choices of sign for $(-1)^{K_g}$ will lead to
different orbifolds.
The $\e(g,h)$, the discrete torsion, is an undetermined phase factor
in the one-loop partition function
when $Q\simeq\ZZ_{N_1}\ldots\ZZ_{N_r}~,r>1$
 \refs{\rIV,\rFIQ}.
By requring factorization and two-loop modular
invariance it has been shown that $\e(g,h)$ has to satisfy \rVT\
\eqn\eTORCOND{
\eqalign{\e(g,h)~~
            &=~~\e(h,g)^{-1}~,  \cr
       \e(h,h)~~&=~~1 ~,  \cr
       \e(g_1g_2,h)~~&=~~\e(g_1,h)\e(g_2,h)~.  \cr}
}
For a $(2,2)$ string vacuum which has an  interpretation in terms of
a string propagating on a \CY\ manifold one has $(-1)^{K_g}=\e(g,h)=1$
for all group elements $g,h\in Q$.

Because the $\tilde X_{6,7}$ are trivial fields and in the ideal,
no monomial $\prod_{\Q_i^h\in\ZZ} (\tilde X_i)^{\ell_i}$ will depend
on $\tilde X_{6,7}$. Thus we only have to study the action on the twisted
 $(c,c)$
vacuum. We now want to argue that after restricting to the physical states,
$\widetilde W(\tilde X_i)/(\tilde jh_1h_2)$  is equivalent to the
theory {\it without} the trivial fields. By using  \eGACT\ and \eTORCOND\
one can show that this is indeed the case  but only if
\eqn\eTORS{
\e(\tilde j|_0,h_i|_0)~=~\e(h_i|_0,h_j|_0)_{i\neq j}~=~-1~,
\qquad (-1)^{K_{h_i|_0}}=-1~,\qquad i,j=1,2~,}
where $\tilde j|_0$, $h_i|_0$ are the actions of $\tilde j$, $h_i$
restricted to the five non-trivial fields. In what follows we will
leave out the $|_0$ unless it is not obvious from the context on which
fields the group element acts.
The first condition follows from \eTORCOND\ ,
the third comes from requiring $(2,2)$ supersymmetry \rIV\
while the second
is chosen so that the trivial fields can be neglected.
A similar analysis for the $(a,c)$ sector gives no more constraints.
Thus we have been able to rewrite the original LG orbifold ${\cal M}=W(X_i)/j$
with $n=7$ to an equivalent theory
$\widetilde {\cal M}\simeq\widetilde W_0(\tilde X_i)/(\tilde j
\cdot h_1\cdot h_2)$
with the torsion  given by \eTORS\ and $n=5$.

Rather than considering the general case for $n=9$, which follows
 straightforwardly from the above, let us look at the following interesting
example.
Let ${\cal M}$ be the LG orbifold constructed from the superpotential $W(X_i)$,
\eqn\eWZ{
W(X_i)~~=~~\sum_{i=1}^9 X_i^3~,}
where the scaling symmetry in the GSO-projection is
$j=(\ZZ_3:1,1,1,1,1,1,1,1,1)~.$
%From a geometric point of view
%$W(x_i)=0$ defines a hypersurface in
%$\cp{(1,1,1,1,1,1,1,1,1)}{8}[3]$.
(For a recent study of the corresponding theory as a hypersurface
in $\CP{8}[3]$ see Ref. \rPLE\ .)
%Although this hypersurface is a K\"ahler manifold the K\"ahler form is not
%a physical state.
A straightforward calculation shows that there are are no $(-1,1)$ states
while the number
of $(1,1)$ states is $84$. Naively one
may think that we cannot apply our technique to get rid of
some of the fields. However, we may as well consider a deformed
superpotential
\eqn\eWZDEF{
\widehat W(X_i)~~=~~\sum_{i=1}^4 (\hat X_i^3 + \hat X_i \hat X_{i+5}^2) +
\hat X_5^3~.}
In fact  both $\widehat {\cal M}=\widehat W/j$ and ${\cal M}$ are
examples of Gepner models. In computing the spectrum and the number
of $E_6$ gauge singlets it has been found that they are the same
for the two theories \rFK\ . This would suggest that they indeed
correspond to the
same conformal field theory.

Define new fields $\tilde X_i$
\eqn\eZREDEF{\tilde X_i~~=~~\hat X_i^{1/2}~,
             \qquad \tilde X_{i+5}~=~\hat X_{i+5} \hat X_i^{1/2}~,
             \qquad i=1,\ldots,4~,}
while $\tilde X_5=\hat X_5$. We obtain the LG orbifold
$\widetilde W(\tilde X_i)/(\tilde j\cdot\ZZ_2^3)$ where the new superpotential
is
\eqn\eWZREDEF{
\widetilde W(\tilde X_i)~=~\widetilde W_0(\tilde X_i) +
            \sum_{i=6}^9 \tilde X_i^2~,\qquad
\widetilde W_0(\tilde X_i)=~\sum_{i=1}^4 \tilde X_i^6 + \tilde X_5^3~,
}
and the field identifications and the GSO-projection $\tilde j$ on the
$\tilde X_i$  are
\eqn\eZIDENT{
\twoeqsalign{h_1~~
            &\define~~(\ZZ_2:1,0,0,0,0,1,0,0,0)~, &
   \qquad    h_2~~&\define~~(\ZZ_2:0,1,0,0,0,0,1,0,0)~,  \cr
       h_3~~&\define~~(\ZZ_2:0,0,1,0,0,0,0,1,0)~, &
   \qquad \tilde j~~&\define~~(\ZZ_6:1,1,1,1,2,3,3,3,3)~.  \cr}
}
Note that $\tilde j\cdot h_1\cdot h_2\cdot h_3\simeq
(\ZZ_2:0,0,0,1,0,0,0,0,1)$ and hence we do not need to consider
the fourth identification. Finally, we obtain
$\widetilde {\cal M}=\widetilde W_0(\tilde X_i)
/(\tilde j\cdot h_1\cdot h_2\cdot h_3)$ with $n=5$
at the expence of introducing non-trivial torsion as in \eTORS\ .
Note that $\widetilde W_0(x_i)=0$ gives a well-defined
\CY\ in $\cp{(1,1,1,1,2)}{4}[6]$. However, our LG orbifold
does not correspond to this model due to the torsion. In fact an
explicit calculation shows that
there are zero $(-1,1)$ and $84$ $(1,1)$ states in $\widetilde {\cal M}$.

\newsec{Generalizations}
Naively one may think that there
are not  many \LG\ models with a potential given by eq, \eWGEN{} or
the corresponding situation for nine fields.
Although any smooth marginal deformation  of $W$ will change the theory
it will not change the spectrum. In particular we will not change the
scaling dimension of the chiral fields $X_i$.  The deformations of $W$ will
take us from one point to another
 in the moduli space of $(1,1)$  deformations, at each point of which
there is a theory ${\cal M}=W/j$.
In a similar fashion,  by using mirror symmetry, we can move in the
moduli space of $(-1,1)$ states of ${\cal M}$ by changing
the superpotential associated to the mirror
${\cal W}$.
By starting from a `special' point we assume that
any other point in the moduli space can be reached by appropriately
deforming the theory.
\footnote{$^4$}{There exist models for which the moduli space cannot
be completely spanned by deforming the superpotential $W$ (or the
mirror potential $W^T$). Our discussion is limited to the
subspace for which the polynomial deformations apply \rGH\ .}
Thus the issue is whether for
every set of weights $k_i$, which admits a polynomial of the
type listed in \rBH\ , it is always possible to find a LG potential of the
form given by \eWGEN{} .
A quick glance at the
list of $c=9$ LG theories \rAR\ shows that this is indeed true
for a large number of models.
However, we will give
examples where the analysis does not apply. But we will argue that
these problems may be circumvented.

In \rIV\ it was shown that a LG orbifold based on a superpotential $W$ of
even degree $\tilde d$ with non-trivial torsion
$\e(j,h)\neq 1$ and
$\det h=-1$  decribes a good string vacuum in the sense of $(2,2)$
supersymmetry.
When $\tilde d$ is odd the $N=2$ world-sheet supersymmetry is broken in the
left hand sector \rIV\ and hence we no longer have a $(2,2)$ vacuum.
 It is by no means  true that
all of the dimensionally reduced models have an even degree superpotential.
In those cases we can not apply our construction. By using mirror symmetry
we will however show that they  correspond to a trivial rewriting of
a \CY\ manifold.

Let ${\cal M}=W(X_i)/j$ where we for simplicity choose $W(X_i)$ to be of the
 form \eWGEN b~. After the field
redefintion we have
\eqn\eWODD{
\widetilde W(\tilde X_i)=\widetilde W_0(\tilde X_i) + \tilde X_6^2 +
\tilde X_7^2~,}
where $\widetilde W_0(\tilde X_i)$ is of odd degree, $\tilde d_0$. Next we
want to construct the mirror of $\widetilde{\cal M}_1=
\widetilde W(\tilde X_i)/(j\cdot\ZZ_2)$. By transposition
\footnote{$^5$}{There is no proof  that
 ${\cal M}$ and its mirror ${\cal W}$ are isomporphic  as conformal field
theories if ${\cal W}$ is obtained from ${\cal M}$ by
transposition with ${\cal M}$ not corresponding to a tensor product of
$N=2$ minimal models.}
 \rBH\ one obtains
\eqn\eWODDTR{
\widetilde W^T(Y_i)=\widetilde W^T_0(Y_i) + Y_6^2 + Y_7^2~.
}
Here $\tilde d_0^T$ is even because the only term in $\widetilde W_0$
in which $\tilde X_4$ appears  is $\tilde X_i\tilde X_4^{2l}$
(or $\tilde X_4^{2l}$) for some $i$
and $l$. After the transposition we have $Y_4^{2l}$ and so the degree
of $\widetilde W^T$ has to be even. Hence  we can apply our construction
to $\widetilde W^T(Y_i)$. We first drop the trivial piece and then divide
by a group $H$ (and the scaling symmetry $j^T$)
such that ${\cal W}_1=\widetilde W_0^T(Y_i)/(j^T\cdot H)$
is the mirror of ${\cal M}_1$. If $\det H=(-1)$ there will be
$\ZZ_2$ torsion. Finally by once again using mirror symmetry we get
$\widetilde {\cal M}=\widetilde W_0(\tilde X_i)/(\tilde j\cdot \tilde H)$
where $\tilde H$ is a quotient group such that $\widetilde {\cal M}$
is the mirror of ${\cal W}_1$.
Since ${\cal W}_1$ is a string vacuum and $\widetilde {\cal M}$ is
an equivalent description of the same SCFT, $\widetilde {\cal M}$ has to be
a string vacuum as well. In fact $\tilde d_0$ is odd so $\widetilde {\cal M}$
is
torsion free and the hypersurface
$\widetilde W(\tilde x_i)=0$ in a weighted $\CP{4}$ describes
a \CY\ . Thus we have shown that ${\cal M}$ is equivalent to a
\CY\ manifold.

Let us illustrate the above analysis with an example.
Let $W(X_i)$ be a degree ten polynomial (with $j=(\ZZ_{10}:2,2,2,2,3,4,5)$)
defined by
\eqn\eWODDEX{
W(X_i)=\sum_{i=1}^4 X_i^5 + X_4 X_6^2 + X_6 X_5^2 + X_7^2~.}
By the fractional transformation, $\tilde X_6=X_6^{1/2},~
\tilde X_5=X_5 X_6^{1/2}$ and $\tilde X_i=X_i$ for the other fields
we obtain
\eqn\eWODDEXFR{
\widetilde W(\tilde X_i)=\widetilde W_0(\tilde X_i)
+ \tilde X_5^2 +\tilde X_7^2~,\qquad
\widetilde W_0=\sum_{i=1}^4 \tilde X_i^5 + \tilde X_4 \tilde X_6^4~,
}
where we need to impose the indentification $h=(\ZZ_2:0,0,0,1,0,1,0)$
to make the field redefinition well-defined. If we continue
along the lines described above we find that the degree for
$\widetilde W_0(\tilde X_i)$
 is odd, $\tilde d_0=5$. Thus $(\det \tilde h|_0)^{\tilde d_0}=-1$ and
$W_0/(\tilde j_0\cdot  h)$ is not
a $(2,2)$ string vacuum. Instead let us consider the mirror of
${\cal M}_1=\widetilde W/
(\tilde j\cdot h)$. By standard arguments \rBH\ the mirror model is
${\cal W}_1=\widetilde W^T(Y_i)/(\tilde j^T\cdot h_1\cdot h_2)$
where $\widetilde W(Y_i)^T$ is the transposed polynomial,
$$
\widetilde W^T(Y_i)=\sum_{i=1}^3 Y_i^5 + Y_4^5 Y_6 + Y_6^4 + Y_5^2 + Y_7^2~,
$$
and the quotient is by $h_1\define(\ZZ_5:4,0,0,1,0,0,0)~,
h_2\define(\ZZ_5:4,0,1,0,0,0,0)$
%
%$$\twoeqsalign{ h_1~~&\define~~(\ZZ_5:4,0,0,1,0,0,0)~, & \qquad
%              h_2~~&\define~~(\ZZ_5:4,0,1,0,0,0,0)~.}$$
%
with $\tilde j^T$ the scaling symmetry of $\widetilde W^T(Y_i)$.
 We can now neglect the trivial piece
{\it without} introducing any torsion; the $h_i$ act trivially on
$Y_{5,7}$. The last step is to use mirror symmetry once more. The mirror
of ${\cal W}_1$ is simply ${\cal M}_2=\widetilde W_0/\tilde j$.
Thus through the successive use of mirror symmetry we have shown that
%$$
%{\cal M}~\fieldred~{\cal M}_1~\mirsym~{\cal W}_1~\fieldred~
%\widetilde {\cal W}_1~\mirsym~{\cal M}_2~.
%$$
${\cal M}$ is equivalent to the \CY\ manifold ${\cal M}_2$.
Note the following. The two trivial terms in
$\widetilde W$ amount to a doubling
of the spectrum in $\widetilde W/\tilde j$ compared to
$\widetilde W_0/\tilde j_0$.
The effect of the $\ZZ_2$ identification $h$
is merely to get rid of the extra copy and hence ${\cal M}_1$ is the
same as  ${\cal M}$. The above analysis reveals the same result.

For some of the Landau-Ginzburg models with $n=7,9$ it is possible to
associate a complete intersection \CY\ manifold
(CICY) \refs{\rR,\rGVW} . A natural question
is how our analysis copes with those theories.
There are two cases which arise. On one hand  models exist for which we
{\it cannot} by any means apply our analysis. An example of such a
theory is \rROLF\
\eqn\eCICYR{
  {\cal M} \in \WK{\CP{3}\cr\CP{2}\cr}{3&1\cr 0&3\cr}~~ :
 ~~\left\{
       \eqaligntwo{
   f(x)   &= ~\sum_{i=1}^4 (x_i)^3       &= ~0~,  \cr
   g(x,y) &= ~\sum_{i=1}^3 x_i y_i^3     &= ~0~,  \cr}\right.}
The matrix notation refers to two polynomial constraints in
$\CP{3}\times
\CP{2}$ where the entries in the first
(second) column are the degrees of the first (second) polynomial.
The corresponding LG orbifold is obtained by adding the polynomial constraints
to form a LG-potential and  to mod out by the appropriate discrete
symmetries as described by Greene, Vafa and Warner \rGVW\ .

The second class of models consists of \LG\ orbifolds
which are equivalent to a CICY in a
product of weighted projective spaces of the form \rRS\
\eqn\eCICYGIB{
  {\cal M} \in
 \WK{\WCP{4}{(k_1,k_1,k_3,k_4,k_5)}\cr
             \CP{1}\hfill}{d_0&k_1\cr 0&2\cr}~~ :
 ~~\left\{
       \eqaligntwo{
   f(x)   &= ~W_0(x_i)      &= ~0~,  \cr
   g(x,y) &= ~\sum_{i=1}^2 x_i y_i^2     &= ~0~,  \cr}\right.}
where $W_0(x_i)$ is a polynomial of degree $d_0$ of the type
 considered in \rBH\ .
Just as above the LG-potential is a sum of the defining polynomials,
\eqn\eWIB{
W(X_i)= W_0(X_i) + \sum_{i=1}^2 X_i Y_i^2~.}
Note that although \eWIB\ is the same as \eWGEN a\ there is an
additional constraint from the vanishing of the first Chern class of ${\cal
M}$,
\eqn\eCHERN{\sum_{i=1}^5 k_i = d_0 + k_1~.}

The idea is to show that $\widetilde {\cal M}$ obtained through
the usual technique of field redefinition and $\ZZ_2$ torsion is equivalent
to a \CY\ defined as a hypersurface in a weighted $\CP{4}$.
 To illustrate this consider a model where $W_0(X_i)$ is
a Fermat \rRS\ ,
\eqn\eWIBEX{
W(X_i)=\sum_{i=1}^5 X_i^4 + X_1 X_6^2 + X_2 X_7^2~.}
Let us define ${\cal M}=W(X_i)/j$. By using fractional transformation,
we construct
$\widetilde {\cal M}=\widetilde W_0(\tilde X_i)
/(\tilde j\cdot  h_1\cdot h_2)$
with the discrete torsion given by \eTORS\ , the $h_i$ as in \eIDENT\ and
\eqn\eWIBEXFR{
\widetilde W(\tilde X_i)=\widetilde W_0(\tilde X_i) +
\tilde X_6^2 + \tilde X_7^2~,\qquad
\widetilde W_0(\tilde X_i)=\tilde X_1^8 + \tilde X_2^8 + \tilde X_3^4 +
\tilde X_4^4 + \tilde X_5^4~,
}
Thus, ${\cal M}\simeq\widetilde {\cal M}$ and we
have found that a perfectly well-defined \CY\ is equivalent to
a \LG\ with torsion. There is, however, no contradiction in this statement.
%as we will now see.
Instead of discarding the trivial terms after the field redefinition we
keep $\widetilde W(\tilde X_i)$. By repeatedly using mirror symmetry
as in the previous examples we find that
$\widetilde W(\tilde X_i)/(\tilde j\cdot  h_1\cdot h_2)\simeq {\cal M}_2$
where ${\cal M}_2= \widetilde W_0(\tilde X_i)/\tilde j$.
Thus we have the following equivalences between the different models,
\eqn\eEQUIV{
\widetilde {\cal M}~~\simeq~~{\cal M}~~\simeq~~{\cal M}_2~.}
Eq. \eEQUIV\ also gives an explanation of the ineffective split
\refs{\rRS,\rTEX}
of ${\cal M}$ to ${\cal M}_2$.
By this it is meant that although the embedding space has been enlarged and
the defining equations changed we have ${\cal M}_2\simeq {\cal M}$.
In terms of the
discussion above the equivalence between the two models is evident.

The last point we would like to address concerns those models for which
we cannot choose a potential of the form given in \eWGEN{} , although we still
restrict to
the class of theories considered in \rBH\ . Since the general
situation is not well understood  we will illustrate the idea by considering
a non-trivial example. However we believe that an argument like
the one below will work for all models.

Let the potential $W(X_i)$ be defined by
\eqn\eWFAIL{
W(X_i)=X_1^3 X_5 + X_2^2X_9 + X_3^2 X_8 +X_4^2X_7+X_5^2X_6+X_6^2X_4+
X_7^2X_3 + X_8^2X_2 + X_9^2 ~,
}
${\cal M}=W(X_i)/j$ (where
$j=(\ZZ_{256}:57,64,80,84,85,86,88,96,128)$) is
 the mirror of the model discussed at the
end of section 2.
$W(X_i)$ together with a polynomial in which $X_2^2X_9$ is replaced
by $X_2^4$ are the only non-degenerate potentials that can be written down
which respects the scaling symmetry $j$. Thus ${\cal M}$ is an example of
a theory for which we cannot carry out our program to full extent---there
is only one term of the form $X_i X_j^2$ and we would need four. But by
using mirror symmetry arguments as in the previous situations the
problem can be  circumvented.

Let ${\cal W}$ be the mirror model of ${\cal M}$,
where ${\cal M}$  is the
GSO-projection of $W^T(Y_i)$
\eqn\eFAILTR{
W^T(Y_i)~=~Y_1^3 + Y_1Y_5^2 + Y_5 Y_6^2 + Y_6 Y_4^2 + Y_4Y_7^2 +
Y_7 Y_3^2 + Y_3 Y_8^2 + Y_8Y_2^2 +Y_2 Y_9^2~.
}
But from section 2 we know that there exists a deformation of $W^T(Y_i)$
given by \eWZDEF\ such that our construction applies. Thus, up
to deformations ${\cal W}$ is
equivalent to $\widetilde {\cal M}=\widetilde W_0(\tilde Y_i)
/(\tilde j\cdot h_1\cdot h_2\cdot h_3)$
with non-trivial discrete torsion as in \eTORS\
%$\e(j,\hat h_i)=\e(\hat h_i,\hat h_j)=-1~,~i=1,2,3$
and the potential
\eqn\eZFR{
\widetilde W_0(\tilde Y_i)=\sum_{i=1}^4 \tilde Y_i^6 + \tilde Y_5^3~,
}
To complete the argument, the mirror of
$\widetilde {\cal M}$ is $\widetilde {\cal W}=
\widetilde W_0(\tilde Y_i)/(\tilde j\cdot \ZZ_6^2\cdot\ZZ_3\cdot\ZZ_2)$
with the actions given by,
$$\twoeqsalign{(\ZZ_6&:5,1,0,0,0,0)~~
                      &~~(\ZZ_6&:5,0,1,0,0,0) \cr
               (\ZZ_3&:2,0,0,0,1)~~
                      &~~(\ZZ_2&:1,0,0,0,0)~.   \cr}
$$
Thus we have shown that ${\cal M}\simeq\widetilde {\cal W}$ where
the latter is an $n=5$ LG orbifold with discrete torsion.

\newsec{Conclusions}
In this paper we have shown that a large class of \LGO\ with more than
five fields can
be rewritten as a SCFT with five fields {\it and} non-trivial discrete
torsion.  Although the construction, which is based on a $\ZZ_2$ field
redefinition, does not apply for all higher dimensional LG
orbifolds we have shown through a number of  examples how this problem
may be solved.
An interesting question is whether it is always possible, not
necessarily by introducing discrete torsion, to rewrite
 any \LGO\ with $n>5$ fields in terms of an $n=5$ model.

Apart from the models which already had an equivalent description as
a \CY\ manifold \rGVW\ and those   for which the degree
of the reduced polynomial is odd
we do not know of a geometrical interpretation for
the theories considered in this paper.
However there are some indications that they may correspond
to three dimensional complex manifolds with vanishing first Chern class.
One can show that the dimensionally reduced LG theory
always has a state of charge $(-3,0)$. This state corresponds to the
 holomorphic $(3,0)$ form in those cases when there is an
equivalent description  in terms of a three dimensional \CY\ manifold \rIV\ .
Hence the existence of the $(-3,0)$ state may suggest that, if there is a
target space picture,  the geometry will be that of a Ricci-flat
complex three-fold.

 We also know that in toroidal compactifications discrete
torsion signals a nontrivial background antisymmetric tensor field
$B_{\m\n}$ \rVT\ . There are solutions of the string equations in which the
dilaton is
not constant for which one is also lead to turn on $B_{\m\n}$ \rS\ .
It has been suggested \rVMS\ that switching on the dilaton as well
as torsion may give a $\s$-model with a three dimensional
(complex) target space. Indeed this is somewhat what
happens in our picture. We hope to return to these questions in
a future publication.

{\bf Acknowledgements}:
The author acknowledges useful discussions with
V.~Batyrev, P.~Candelas, X.~de la Ossa,
T.~H\"ubsch, M.~Kreuzer, J.~Louis, F.~Quevedo and R.~Schimmrigk. This work was
supported by  the America-Scandinavian Foundation, the Fulbright Program, the
NSF grants PHY 8904035, 9009850 and the Robert~A.~Welch foundation.
The author also would like to thank the
Theory Division at CERN for hospitality where this work was started.

\vfill \eject

\listrefs

\bye